# Passive state preparation for quantum key distribution with phase encoding

Roman Shakhovoy
QRate
Moscow, Russia
r.shakhovoy@goqrate.com

*Abstract*—We propose here a method of passive state preparation for quantum key distribution with phase encoding based on the measurement of the phase difference between pulses of a gain-switched laser. The features of the optical scheme of the transmitter are discussed and the results of computer simulation are presented to demonstrate the effectiveness of the proposed method.

## I. Introduction

An important challenge in the field of quantum key distribution (QKD) today is the creation of low-cost, compact devices for metropolitan area QKD networks. One of the solutions in this direction is a passive transmitter, where quantum states are prepared without modulators [1, 2] and where randomized light sources (lasers in a gain switching mode or even thermal sources [3]) are generally used as an entropy source. Such transmitters have a number of advantages. First, they potentially simplify the schematic, which reduces the cost of the transmitter. In addition, the absence of active modulators allows mitigating leaks associated with them; in particular, a Trojan horse attack turns out to be irrelevant here. Finally, passive preparation does not require a random number generator (RNG), which makes the system much simpler and removes the limitation on the rate of quantum state generation caused by the finite rate of generation of random bits.

In this work, we propose a method of passive state preparation for the BB84 protocol [4] with phase encoding. The main idea is to use a gain-switched laser performing measurement of the phase difference between adjacent laser pulses. Additionally, we use optical injection to create a time delay between pairs of laser pulses.

Section II discusses the optical design of the transmitter, Section III describes the mathematical model of the optically coupled laser system, Section IV shows simulations demonstrating the proposed passive state preparation method, and Section V discusses the results obtained.

## II. General description of a scheme

A simplified schematic of a transmitter (Alice) implementing passive state preparation for QKD with phase encoding is shown in Fig. 1. The transmitter uses optically coupled lasers connected through an optical circulator (OC). Both lasers operate in a gain switching mode, and the pulse repetition period of the slave laser, $\Delta T$, is three times less than the pulse repetition period of the master laser, so that phase locking occurs for only $1/3$ of the slave laser pulses.

An optical filter is installed at the output port of the circulator. Its central transmission wavelength corresponds to the wavelength of the slave laser. The detuning of the master and slave lasers is chosen such that it exceeds the bandwidth of the WDM filter, so that radiation at the master wavelength is blocked by a filter.

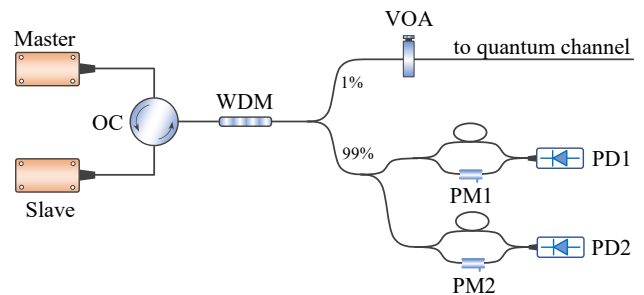

Fig. 1. A simplified schematic of a transmitter implementing passive state preparation for QKD with phase encoding: OC is an optical circulator, PM1, PM2 are phase modulators, PD1, PD2 are photodetectors, VOA is a variable optical attenuator, WDM is an optical bandpass filter.

When the master emits an optical pulse, the radiation in the corresponding pulse of the slave laser changes its wavelength due to the locking effect, which makes the slave's wavelength equal to the wavelength of the master laser. Thus, only those slave's pulses that were generated in the absence of master radiation pass through the optical filter. As a result, a pulse sequence consisting of separate pairs of pulses appears at the output of the filter. Each pair corresponds to a certain quantum state, useful information about which is contained in the phase difference between the pulses. Since the time delay between pulses in a pair is equal to $\Delta T$, and the delay between pairs is equal to $2\Delta T$, the frequency of preparation of quantum states is $f = 1/(3\Delta T)$. After the filter, a small part of radiation is diverted into a variable optical attenuator, where the laser pulses are attenuated and then sent into the quantum channel.

Since the slave laser operates in a gain switching mode, the phase differences between the pulses in different pairs will be random, so Alice needs to measure them to know what states she is sending to Bob. To do this, most of the radiation is sent to a phase detector (see Fig. 1), where Alice measures the phase difference between adjacent pulses. The phase detector in our scheme is represented by two unbalanced Mach-Zehnder interferometers, the delay line in which corresponds to the pulse repetition period of the slave laser. The results of the interference (intensities of interfering pulses), measured at the outputs of interferometers with photodetectors, are used to determine the phase difference between the pulses (see section IV for details).

Since here we are only interested in the method of preparing quantum states, we omit the features of the receiver. Note only that to decode the states sent by Alice, Bob also needs to measure the result of the interference of neighboring pulses, i.e. its optical scheme must contain an unbalanced interferometer.

### III. LASER RATE EQUATIONS

In most cases, to describe the dynamics of semiconductor lasers in the presence of optical injection, it is sufficient to use the model of rate equations [5, 6]. If the pumping of the master and slave lasers is a function of time, then the dynamics of lasing is described by a system of six nonlinear differential equations, three of which describe the time dependence of the normalized intensity (number of photons) $Q^M$, the number of carriers $N^M$ and the phase $\varphi^M$ of master:

$$\frac{dN^M}{dt} = \frac{I^M}{e} - \frac{N^M}{\tau_e^M} - \frac{Q^M}{\Gamma^M \tau_{ph}^M} G^M + F_N^M,$$

$$\frac{dQ^M}{dt} = (G^M - 1)\frac{Q^M}{\tau_{ph}^M} + C_{sp}^M \frac{N^M}{\tau_e^M} + F_Q^M, \quad (1)$$

$$\frac{d\varphi^M}{dt} = \frac{\alpha^M}{2\tau_{ph}^M}(G_L^M - 1) + F_\varphi^M,$$

and the other three equations determine the dynamics of the number of photons $Q$, the number of carriers $N$ and the phase $\varphi$ of the slave laser:

$$\frac{dN}{dt} = \frac{I}{e} - \frac{N}{\tau_e} - \frac{Q}{\Gamma \tau_{ph}} G + F_N,$$

$$\frac{dQ}{dt} = (G-1)\frac{Q}{\tau_{ph}} + C_{sp}\frac{N}{\tau_e} +$$
$$+ 2\kappa_{inj}\sqrt{Q^M Q}\cos(\Delta\omega_{inj}t + \varphi^M - \varphi) + F_Q, \quad (2)$$

$$\frac{d\varphi}{dt} = \frac{\alpha}{2\tau_{ph}}(G_L - 1) +$$
$$+ \kappa_{inj}\sqrt{\frac{Q^M}{Q}}\sin(\Delta\omega_{inj}t + \varphi^M - \varphi) + F_\varphi.$$

In equations (1) and (2), the superscript $M$ indicates the quantities corresponding to the master laser. The gain $G$ here is defined as a dimensionless normalized quantity as follows: $G = (N - N_{tr})/(N_{th} - N_{tr})$, where $N_{tr}$ is the number of carriers at transparency, and $N_{th}$ is the threshold number of carriers. Onwards, $\tau_e$ and $\tau_{ph}$ are the lifetimes of carriers and photons, respectively; $\Gamma$ is the confinement factor; $C_{sp}$ is the fraction of spontaneously emitted photons falling into the laser mode under consideration; $\alpha$ is the linewidth enhancement factor (the so-called Henry factor); $I$ is the pump current; $e$ is the absolute value of the electron charge; $\kappa_{inj}$ is the coupling coefficient between the control and slave lasers, which determines the efficiency of optical injection; $\Delta\omega_{inj}$ is the frequency detuning of lasers and, finally, $F_N$, $F_Q$, and $F_\varphi$ are random Langevin forces, responsible for fluctuations in the number of carriers, number of photons and phase, respectively. Langevin forces are written explicitly as follows (in the form of differentials):

$$F_Q dt = 2\sqrt{\frac{C_{sp}NQ}{2\tau_e}}\left(\cos\varphi\, dW^A + \sin\varphi\, dW^B\right),$$

$$F_\varphi dt = \sqrt{\frac{C_{sp}N}{2\tau_e Q}}\left(\cos\varphi\, dW^B - \sin\varphi\, dW^A\right),$$

$$F_N dt = -2\sqrt{\frac{C_{sp}NQ}{2\tau_e}}\left(\cos\varphi\, dW^A + \sin\varphi\, dW^B\right) + \quad (3)$$
$$+ \sqrt{\frac{2N}{\tau_e}}dW^C,$$

where $W^A$, $W^B$, and $W^C$ are independent Wiener processes. It should be noted that for the master laser in equations (3) a superscript $M$ should be assigned to each parameter; in addition, it is necessary to introduce three other independent Wiener processes for the master (thus, there should be 6 of them in total).

### IV. SIMULATIONS

The optical pulse trains emerging from the master and slave lasers (before they pass through the optical bandpass filter) are shown in Fig. 2(a). To simulate optical signals, we used rate equations (1)–(2) with Langevin forces from (3). It is assumed that both lasers operate in gain-switching mode and generate pulse sequences with different repetition periods: the master's pulse repetition period is three times longer than the slave laser's pulse repetition period. The signal after the optical bandpass filter is shown in Fig. 2(b), where the pulse sequence is divided into pairs of pulses with a random phase difference between them. (A second order Butterworth filter was used to simulate optical filtering.) To explicitly show that the generated states are unknown before measurement, we use for these states the designation $B_r$, where $B$ can take values $X$, $Y$, and the subscript $r$ can take values 0 or 1.

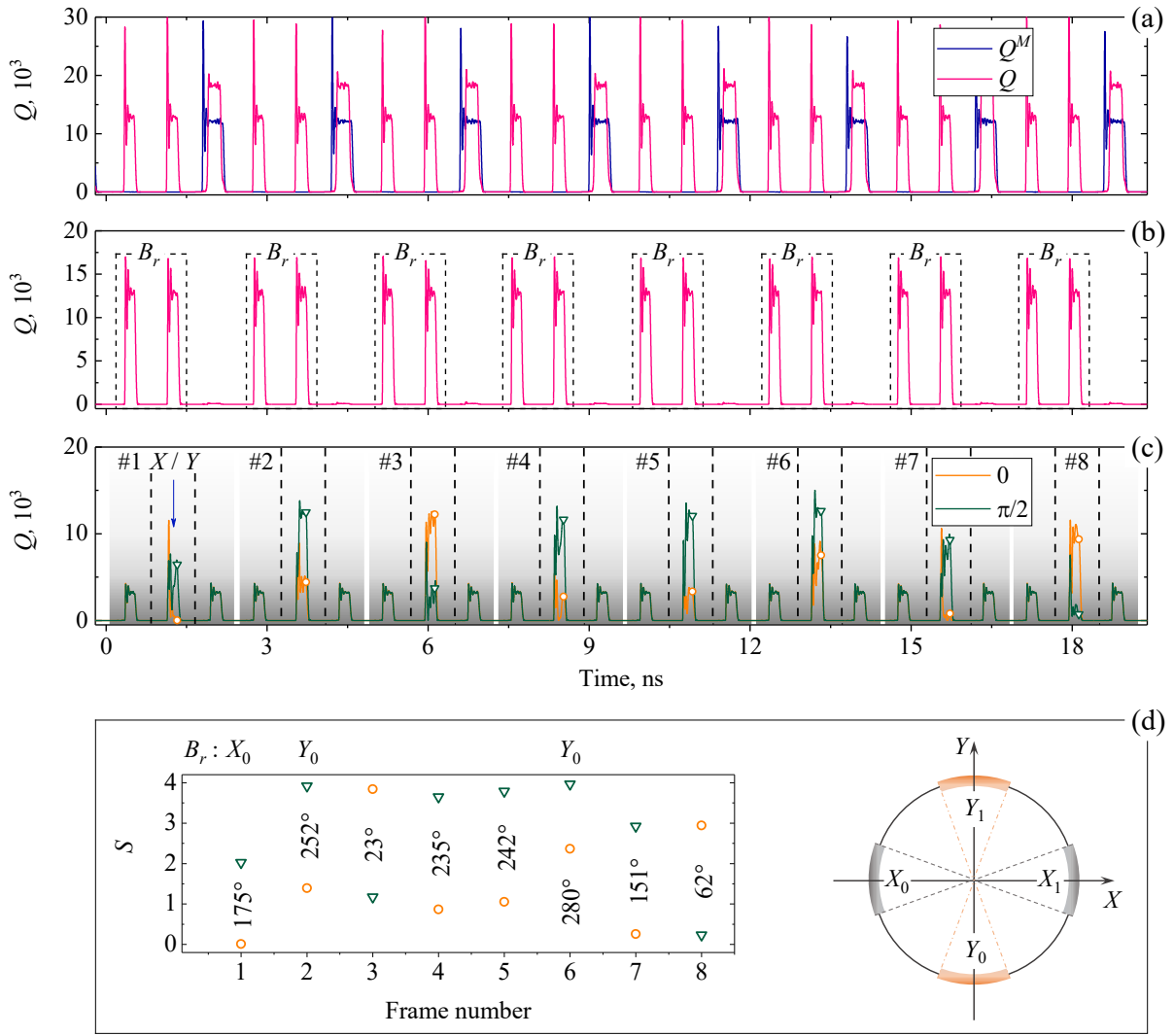

Fig. 2. (a) Simulations of optical pulses from master and slave lasers. (b) Pulse sequence of the slave laser after optical filtering. (c) Results of the pulse interference in Alice's interferometers with additional phase incursions 0 and $\pi/2$. (d) Phase differences between pulses obtained using formulas (4)–(6) (on the left) and a complex plane with sectors limiting the ranges of angles corresponding to different quantum states (on the right).

As discussed in Section II, Alice needs to measure the phase difference for each pair of pulses. To do this, she uses a system of two interferometers shown in Fig. 1, which act as a phase detector. From the experimental point of view, before measuring the interference of pulses with such a system, it is necessary to first adjust the phase modulators PM1 and PM2 by setting phases (or rather voltages) on them such that the phase differences between the interfering pulses in the two interferometers differ by $\pi/2$. This means that if the phase difference between interfering pulses in one interferometer is $\varphi_r$, then the same pulses must interfere with a phase difference $\varphi_r + \pi/2$ in the second interferometer. Simulations of the interference results in two interferometers configured in this way are shown in Fig. 2(c). Here the sequence of pulses is divided into frames, each containing three time slots (for clarity, we will call these slots "early", "central" and "late"). The intensity values of the resulting pulses for the first interferometer are shown by empty circles (for definiteness, we will assume that the phase modulator of this interferometer introduces phase 0), and for the second interferometer, the phase modulator of which introduces an additional phase $\pi/2$, is shown by empty triangles.

From an experimental point of view, measuring the interference result on Alice's side corresponds to measuring the pulse intensity in the central time slot and then normalizing it to the pulse intensity in the early or late time slot. If the normalized value $S$ of the interference signal is around 4 (naturally, with some error due to the imperfection of the measuring equipment), then the interference is constructive; if this value is around 0, then it is destructive. It should be noted here that to measure the intensity of interfering pulses with sufficient accuracy, it is advisable to digitize the photodetector signal using an analog-to-digital converter (ADC) that has the largest possible bandwidth; narrowband ADCs will significantly distort the pulse shape and will not allow accurate measurement of the normalized signal $S$.

In Fig. 2(d) (on the left) normalized values of the intensities of interfering pulses from different frames are shown using the same notations as in Fig. 2(c) (i.e. with empty circles and triangles). For definiteness, we will use the notation $S_1$ for the normalized interference signal corresponding to the phase shift 0, and $S_2$ for the normalized interference signal with an additional phase shift $\pi/2$. Given the values of the normalized interference signals $S_1$ and $S_2$ for a given pair of pulses, the phase difference between them can be determined using the following formulas:

$$\theta_{1,2} = \pm \arccos\left(\frac{S_1 - 2}{2}\right), \qquad (4)$$

$$\varphi_1 = \pi - \arcsin\left(1 - \frac{S_2}{2}\right), \qquad (5)$$

$$\varphi_2 = \arcsin\left(1 - \frac{S_2}{2}\right), \qquad (6)$$

where $\theta_i$ and $\varphi_i$ correspond to possible values of the desired phase difference. The true value of the phase difference corresponds to the value of $\theta_i$ that coincides with one of $\varphi_i$ values. For example, if $\theta_1 = \varphi_1$ or $\theta_1 = \varphi_2$ (in this case, the inequalities $\theta_2 \neq \varphi_1$ and $\theta_2 \neq \varphi_2$ must be satisfied, respectively), then the true value of the phase difference is equal to $\theta_1$. (The values of $\theta_i$ и $\varphi_i$ must be limited to the range from 0 to $2\pi$.) It is important to keep in mind, however, that in a real experiment, all four angle values calculated using formulas (4)–(6) may turn out to be different due to noise in the measuring instruments and in the laser itself. Therefore, from a practical point of view, the correct value corresponds to searching for the minimum element of the sequence $\{\theta_1 - \varphi_1, \theta_1 - \varphi_2, \theta_2 - \varphi_1, \theta_2 - \varphi_2\}$. Thus, if the element $\theta_2 - \varphi_1$ takes the minimum value, then the phase difference between the pulses should be set equal to $\theta_2$.

The values of the phase difference between different pairs of pulses, determined by the method just described, are indicated near the corresponding values of the normalized signals in Fig. 2(d) on the left. Since the described method allows Alice to distinguish states with phase differences $\pi/2$ and $3\pi/2$, she can use two non-orthogonal "phase" bases: $X$-basis that uses phase differences 0 and $\pi$ for states $X_1$ и $X_0$, and $Y$-basis that uses the phase differences $\pi/2$ и $3\pi/2$ for the states $Y_1$ and $Y_0$. However, with such a choice of angles, Alice will have to discard most of the states, so it is advisable to use arcs rather than points on the complex plane as a signal constellation, as shown in Fig. 2(d) on the right. Here, for definiteness, we have chosen the central angles of all arcs to be equal to $40°$. Figure 2(d) on the left shows that with this choice of angle range, three of the eight frames can be assigned certain states: for the frames ##2,6 and for the frame #1.

## V. DISCUSSION

It is worth noting that the time delay between pairs of pulses in Fig. 2(b) is not mandatory. Moreover, in the absence of such a delay, the frequency of preparing states will automatically increase. The interference pattern, however, will change in this case and will represent a sequence of pulses with random amplitudes, which should be divided into frames of two time slots. One then should work only with even or odd (depending on the reference point) pulses. From an experimental point of view, this situation is much less convenient than that shown in Fig. 2(c), where the central time slots are always separated by pulses with a fixed amplitude. Indeed, the latter are convenient to use for signal normalization, which is necessary to determine the phase difference using formulas (4)–(6). In addition, in the absence of a time delay between neighboring states, the phase differences between the pulses turn out to be related, since pulses from neighboring states participate in the interference. Strictly speaking, in this case we cannot talk about phase randomization, since Eve can probably obtain information about the phases of neighboring states. Although we have not studied this issue in detail, at first glance it seems that in this case Eve may try to carry out an attack like a sequential attack [7] on the differential phase shift QKD protocol [8]. For these reasons, the use of a time delay between adjacent pairs of pulses seems reasonable to us.

Another feature of the scheme we propose is the use of optically coupled lasers to form "empty" time slots. In principle, the pulse sequence in Fig. 2(b) can be created with a solitary laser using an appropriate electrical signal for pumping. However, in this case, the shape of adjacent pulses may differ significantly, especially if the bias current is significantly below the threshold [6]. This may negatively affect the visibility of interference, which will lead to an increase in the quantum bit error rate, and also will not allow Alice to accurately determine the phase difference between pulses.

The difference in shape of pulses in an irregular pulse sequence is associated with the finite lifetime of carriers, or more precisely with the fact that the evolution of the number of carriers between pulses belonging to the same pair will differ from the evolution of $N$ between pulses belonging to different pairs. To reduce this effect, the bias current is usually set above the threshold. This solution, however, is not suitable for the considered method of passive state preparation since it violates the randomization of pulse phases [9]. The proposed method of optical injection followed by optical filtering, as demonstrated by our simulations, does not lead to distortion of the pulse shape since both lasers generate regular pulse sequences.

We also note the possibility of using another method of driving lasers in Fig. 1, which can further enhance interference visibility and reduce the impact of jitter. The master can create a pulse sequence similar to that shown in Fig. 2(b), i.e. generate pairs of pulses with a time delay of $2\Delta T$ between pairs and thus cover the 2/3 of pulses of the slave laser. The central transmission wavelength of the optical filter in this case must coincide with the master wavelength, which will allow filtering every third pulse in the slave laser signal. As a result, we will again obtain the sequence shown in Fig. 2(b), however, now the optical pulses will be less chirped and will exhibit lower jitter, which will have a positive effect on their interference. Of course, as mentioned above, the master pulses with such pumping may differ in shape, which again can negatively affect the interference of the slave laser pulses. This is especially true when this method is used at high pulse repetition rates (more than 1 GHz). At low frequencies, however, the master pulses can be made quite long, whereas the slave laser pulses can be made short, which will neutralize the influence of the master pulse shape on the slave laser pulse shape. So, this alternative method of preparing states can be useful at low pulse repetition rates.

## VI. Conlusions

The method we propose for the passive preparation of quantum states has high potential for creating devices for metropolitan area QKD networks. Note, however, that today it has limited practical value due to the need in an integral-optical phase detector. The cost of the latter today may exceed the cost of standard modulators based on lithium niobate, which negates the advantages of this approach (for purely economic reasons) over traditional QKD schemes with active preparation of states. However, with the further development and reduction in cost of photonic integrated schemes, the value of our method will increase significantly. In addition, our approach can be used not only for phase, but also for amplitude-phase modulation, which significantly expands the possibilities for preparing quantum states.

# Пассивное приготовление состояний для квантового распределения ключей с фазовым кодированием

Шаховой Роман

КуРэйт, Москва, Россия

r.shakhovoy@goqrate.com

*Аннотация*—В данной работе предлагается метод пассивного приготовления состояний для квантового распределения ключей с фазовым кодированием, основанный на измерении разности фаз между импульсами лазера режиме переключения усиления. Обсуждаются особенности оптической схемы передатчика. Для демонстрации эффективности предлагаемого метода приведены результаты компьютерного моделирования.

*Ключевые слова—оптическая инжекция; квантовое распределение ключей; пассивное приготовление состояний*

## VII. Введение

Важной задачей в области квантового распределения ключей (КРК) сегодня является создание недорогих компактных устройств для городских сетей. Одним из направлений развития систем КРК в рамках решения этой задачи являются пассивные передатчики, в которых для приготовления квантовых состояний применяются оптические схемы без модуляторов [1, 2], а в качестве источника энтропии используются рандомизированные источники света, например, лазеры в режиме переключения усиления или даже тепловые источники [3]. Такие передатчики имеют ряд преимуществ. Прежде всего, они потенциально позволяют упростить схему, что уменьшает стоимость передатчика. Кроме того, отсутствие активных модуляторов позволяет нивелировать связанные с ними утечки, в частности атака троянским конем оказывается здесь нерелевантной. Наконец, при пассивном приготовлении не требуется генератор случайных чисел (ГСЧ), что делает систему значительно проще, а также снимает ограничение на скорость генерации квантовых состояний, обусловленное конечной скоростью генерации случайных бит.

В данной работе мы предлагаем способ пассивного приготовления состояний для протокола BB84 [4] с фазовым кодированием. Основная идея заключается в измерении разности фаз между соседними лазерными импульсами лазера в режиме переключения усиления и последующей селекцией состояний. Дополнительно мы используем оптическую инжекцию, чтобы создать временную задержку между парами лазерных импульсов, кодирующих квантовое состояние.

В разделе II обсуждается оптическая схема передатчика, в разделе III описана математическая модель используемой нами системы оптически связанных лазеров, в разделе IV показаны симуляции, демонстрирующие предлагаемый нами способ пассивного приготовления состояний, а в разделе V обсуждаются полученные результаты.

## VIII. Общее описание схемы

Упрощенная схема передатчика (Алисы) с пассивным приготовлением состояний для КРК с фазовым кодированием показана на Рис. 1. В схеме используются оптически связанные лазеры, соединенные через оптический циркулятор (ОЦ). Оба лазера работают в режиме переключения усиления, причем период следования импульсов ведомого лазера, $\Delta T$, втрое меньше периода следования импульсов мастера, так что фазовая синхронизация возникает только для 1/3 импульсов ведомого лазера. На выходе из циркулятора установлен оптический фильтр, центральная длина волны пропускания которого соответствует длине волны ведомого лазера. Отстройка мастера и ведомого лазеров по длине волны выбрана такой, чтобы она превышала полосу пропускания WDM фильтра, т.е. чтобы излучение на длине волны мастера блокировалось фильтром. Когда мастер испускает оптический импульс, излучение в соответствующем импульсе ведомого лазера меняет длину волны, которая за счет эффекта захвата частоты становится равной длине волны мастера. Таким образом, через оптический фильтр будут проходить только те импульсы ведомого лазера, которые генерировались в отсутствие излучения мастера. В результате на выходе из фильтра возникает импульсная последовательность, состоящая из отдельных пар импульсов. Каждая пара соответствует некоторому квантовому состоянию, полезная информация о котором содержится в разности фаз между импульсами. Поскольку временная задержка между импульсами в паре равна $\Delta T$, а задержка между парами равна $2\Delta T$, то частота приготовления квантовых состояний $f = 1/(3\Delta T)$. После фильтра небольшая часть излучения отводится в регулируемый оптический аттенюатор, где лазерные импульсы ослабляются, после чего отправляются в квантовый канал.

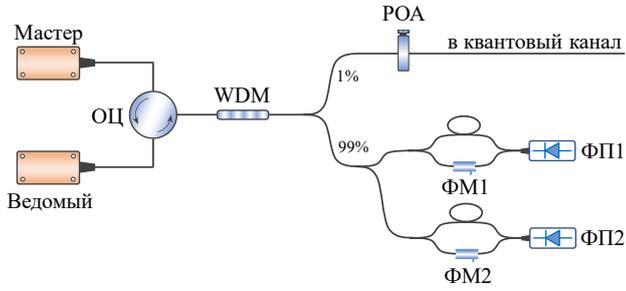

Рис. 3. Упрощенная схема передатчика (Алисы) с пассивным приготовлением состояний для КРК с фазовым кодированием: ОЦ – оптический циркулятор, ФМ1, ФМ2 – фазовые модуляторы, ФП1, ФП2 – фотоприемники, РОА – регулируемый оптический аттенюатор, WDM – полосовой оптический фильтр.

Поскольку ведомый лазер работает в режиме переключения усиления, разности фаз между импульсами в разных парах будут случайными, поэтому Алисе необходимо их предварительно измерить, чтобы знать, какие состояния она отправляет Бобу. Для этого бо́льшая часть излучения отправляется в фазовый детектор (см. Рис. 1), где Алиса измеряет разницу фаз между соседними импульсами. Фазовый детектор выполнен в виде двух несбалансированных интерферометров Маха-Цендера, линия задержки в которых соответствует периоду следования импульсов ведомого лазера. Результаты интерференции (интенсивности интерферирующих импульсов), измеряемые на выходах интерферометров с помощью фотоприемников, используются для определения разности фаз между импульсами (см. раздел IV).

Поскольку в данной работе нас интересует только метод приготовления квантовых состояний, мы оставляем без внимания особенности работы приемника. Отметим только, что для декодирования отправляемых Алисой состояний Бобу также необходимо измерять результат интерференции соседних импульсов, т.е. в его оптической схеме должен присутствовать несбалансированный интерферометр. Выбор базиса на стороне Боба, как обычно, может осуществляться изменением фазы на фазовом модуляторе в одном из плеч интерферометра. Важно, чтобы полоса фазового модулятора Боба была достаточной для переключения фазы с частотой следования лазерных импульсов. Заметим, что это требование не относится к фазовым модуляторам в интерферометрах Алисы, которые, как будет ясно ниже, должны лишь задавать постоянную фазу.

## IX. Лазерные скоростные уравнения

В большинстве случаев для описания динамики полупроводниковых лазеров при наличии оптической инжекции достаточно использовать приближение скоростных уравнений [5, 6]. Если накачка управляющего и управляемого лазеров является функцией времени, то динамика лазерной генерации описывается системой из шести нелинейных дифференциальных уравнений, три из которых описывают зависимость от времени нормированной интенсивности (числа фотонов) $Q^M$, числа носителей $N^M$ и фазы $\varphi^M$ мастера:

$$\frac{dN^M}{dt} = \frac{I^M}{e} - \frac{N^M}{\tau_e^M} - \frac{Q^M}{\Gamma^M \tau_{ph}^M} G^M + F_N^M,$$
$$\frac{dQ^M}{dt} = \left(G^M - 1\right)\frac{Q^M}{\tau_{ph}^M} + C_{sp}^M \frac{N^M}{\tau_e^M} + F_Q^M, \quad (7)$$
$$\frac{d\varphi^M}{dt} = \frac{\alpha^M}{2\tau_{ph}^M}\left(G_L^M - 1\right) + F_\varphi^M,$$

а другие три определяют динамику числа фотонов $Q$, числа носителей $N$ и фазы $\varphi$ ведомого лазера:

$$\frac{dN}{dt} = \frac{I}{e} - \frac{N}{\tau_e} - \frac{Q}{\Gamma \tau_{ph}} G + F_N,$$
$$\frac{dQ}{dt} = (G-1)\frac{Q}{\tau_{ph}} + C_{sp}\frac{N}{\tau_e} +$$
$$+ 2\kappa_{inj}\sqrt{Q^M Q}\cos(\Delta\omega_{inj} t + \varphi^M - \varphi) + F_Q, \quad (8)$$
$$\frac{d\varphi}{dt} = \frac{\alpha}{2\tau_{ph}}(G_L - 1) +$$
$$+ \kappa_{inj}\sqrt{\frac{Q^M}{Q}}\sin(\Delta\omega_{inj} t + \varphi^M - \varphi) + F_\varphi.$$

В уравнениях (1) и (2) верхний индекс $M$ указывает на величины, соответствующие мастеру. Усиление $G$ здесь определено в виде безразмерной нормированной величины следующим образом: $G = (N - N_{tr})/(N_{th} - N_{tr})$, где $N_{tr}$ – число носителей, при котором материал активного слоя прозрачен на длине волны рассматриваемой лазерной моды, а $N_{th}$ – пороговое число носителей. Далее $\tau_e$ и $\tau_{ph}$ – времена жизни носителей и фотонов, соответственно; Г – коэффициент удержания моды (фактор конфайнмента); $C_{sp}$ – средняя доля спонтанно излученных фотонов, попадающих в рассматриваемую лазерную моду; $\alpha$ – коэффициент уширения линии (так называемый фактор Генри); $I$ – ток накачки; $e$ – абсолютное значение заряда электрона; $\kappa_{inj}$ – коэффициент связи между управляющим и ведомым лазерами, определяющий эффективность оптической инжекции; $\Delta\omega_{inj}$ – расстройка лазеров по частоте и, наконец, $F_N$, $F_Q$ и $F_\varphi$ – случайные ланжевеновские силы, отвечающие за флуктуации числа носителей, числа фотонов и фазы, соответственно. В явном виде ланжевеновские силы записываются следующим образом (в форме дифференциалов):

$$F_Q dt = 2\sqrt{\frac{C_{sp}NQ}{2\tau_e}}\left(\cos\varphi\, dW^A + \sin\varphi\, dW^B\right),$$

$$F_\varphi dt = \sqrt{\frac{C_{sp}N}{2\tau_e Q}}\left(\cos\varphi\, dW^B - \sin\varphi\, dW^A\right),$$

$$F_N dt = -2\sqrt{\frac{C_{sp}NQ}{2\tau_e}}\left(\cos\varphi\, dW^A + \sin\varphi\, dW^B\right) +$$

$$+ \sqrt{\frac{2N}{\tau_e}} dW^C,\qquad(9)$$

где $W^A$, $W^B$ и $W^C$ – независимые винеровские процессы. Следует отметить, что для мастера в уравнениях (3) каждой величине необходимо приписать индекс $M$; кроме того, для него необходимо ввести три других независимых винеровских процесса (таким образом, всего их должно быть 6).

## X. СИМУЛЯЦИИ

Последовательности оптических импульсов, выходящие из мастера и ведомого лазеров, показанных на Рис. 1 (до прохождения ими оптического полосового фильтра) показаны на Рис. 4(а). Для симуляций оптических сигналов использовались системы уравнений (1)–(2) с ланжевеновскими шумами (3). Предполагается, что оба лазера работают в режиме переключения усиления и генерируют последовательности импульсов с разными периодами следования: период следования импульсов мастера в три раза превышает период следования импульсов ведомого лазера. Сигнал после оптического полосового фильтра показана на Рис. 4(б), где последовательность импульсов разбита на пары импульсов со случайной разностью фаз между ними. (Для симуляции оптической фильтрации использовалась фильтр Баттерворта 2-го порядка.) Чтобы явно показать, что генерируемые состояния до измерения не известны, мы используем на Рис. 4(б) для этих состояний обозначение $B_r$, где $B$ может принимать значения $X$, $Y$, а индекс $r$ может принимать значения 0 или 1.

Как указывалось в разделе II, Алисе необходимо измерить разность фаз для каждой пары импульсов. Для этого она использует систему из двух интерферометров, показанных на Рис. 1, которые выполняют роль фазового детектора. Перед измерением интерференции импульсов в данной схеме необходимо вначале провести настройку фазовых модуляторов ФМ1 и ФМ2, установив на них такие напряжения, чтобы разности фаз между интерферирующими импульсами в двух интерферометрах отличались на $\pi/2$. Это означает, что если разность фаз между интерферирующими импульсами в одном интерферометре равна $\varphi_r$, то во втором интерферометре эти же импульсы должны интерферировать с разностью фаз $\varphi_r + \pi/2$. Симуляции результатов интерференции в двух интерферометрах, настроенных таким способом, показаны на Рис. 4(в). Здесь последовательность импульсов разбита на фреймы по три временных слота (будем эти слоты для определенности называть «ранний», «центральный» и «поздний»). Как понятно из Рис. 4(в), результат интерференции находится в центральном слоте. Значения интенсивности результирующих импульсов для верхнего интерферометра показаны пустыми кружка́ми (будем для определенности считать, что фазовый модулятор этого интерферометра вносит фазу 0), а для нижнего интерферометра, фазовый модулятор которого вносит дополнительную фазу $\pi/2$, – пустыми треугольниками.

С экспериментальной точки зрения, измерение результата интерференции на стороне Алисы соответствует измерению интенсивности импульса во центральном временном слоте с последующей его нормировкой на интенсивность импульса в раннем или позднем временном слоте. Если нормированное значение $S$ сигнала интерференции равно 4 (естественно, с некоторой погрешностью, обусловленной несовершенством измерительного оборудования), то интерференция является конструктивной, если это значение равно 0 – деструктивной. Здесь следует отметить, что для более точного измерения результата интерференции желательно проводить оцифровку сигнала фотоприемника с помощью аналого-цифрового преобразователя (АЦП), имеющего как можно бо́льшую полосу пропускания; узкополосные АЦП будут существенно искажать форму импульса и не позволят достаточно точно измерить значение нормированного сигнала $S$.

На Рис. 4(г) слева для нормированных значений результатов интерференции, полученных в разных фреймах, используются такие же обозначения, как и на Рис. 4(в) (т.е. пустые кружки́ и треугольники). Для определенности будем использовать обозначение $S_1$ для нормированного интерференционного сигнала,

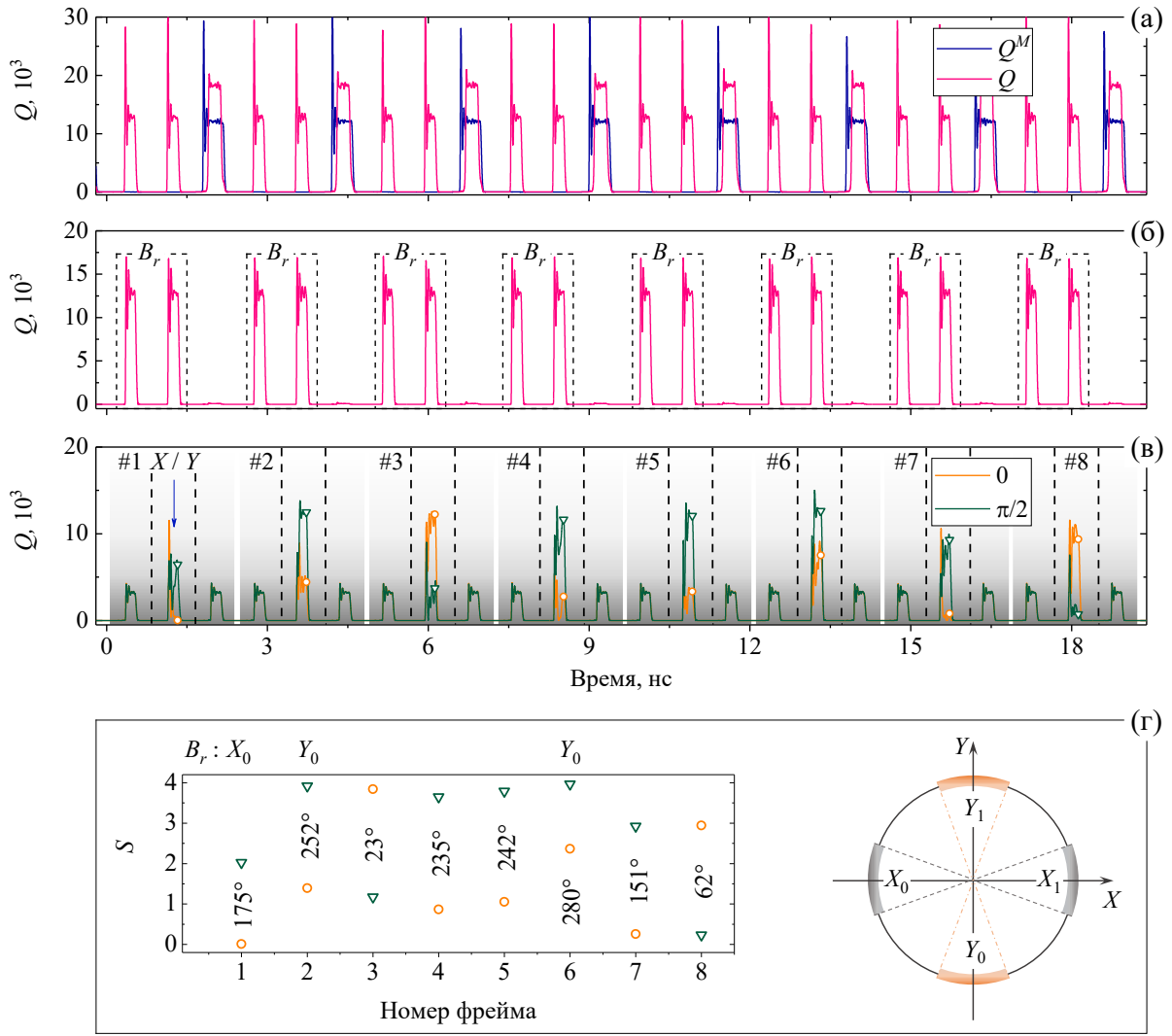

Рис. 4. (а) Симуляции оптических импульсов мастера и ведомого лазера. (б) Последовательность импульсов ведомого лазера после оптической фильтрации. (в) Результаты интерференции импульсов в интерферометрах с дополнительными набегами фаз 0 и $\pi/2$. (г) Полученные по формулам (4)–(6) разности фаз между импульсами (слева) и комплексная плоскость с секторами, ограничивающими диапазоны углов, соответствующих разным квантовым состояниям.

соответствующего набегу фазы 0, и $S_2$ – для нормированного интерференционного сигнала с дополнительным набегом фазы $\pi/2$. Имея значения нормированных интерференционных сигналов $S_1$ и $S_2$ для данной пары импульсов, разность фаз между ними можно определить, используя следующие формулы:

$$\theta_{1,2} = \pm\arccos\left(\frac{S_1-2}{2}\right), \quad (10)$$

$$\varphi_1 = \pi - \arcsin\left(1-\frac{S_2}{2}\right), \quad (11)$$

$$\varphi_2 = \arcsin\left(1-\frac{S_2}{2}\right), \quad (12)$$

где $\theta_i$ и $\varphi_i$ соответствуют возможным значениям искомой разности фаз. Истинная величина разности фаз соответствует тому значению $\theta_i$, которое совпадает с одним из значений $\varphi_i$. Например, если $\theta_1 = \varphi_1$ или $\theta_1 = \varphi_2$ (при этом должны, соответственно, выполняться неравенства $\theta_2 \neq \varphi_1$ и $\theta_2 \neq \varphi_2$), то истинное значение разности фаз равно $\theta_1$. (Значения $\theta_i$ и $\varphi_i$ необходимо ограничивать интервалом от 0 до $2\pi$.) Важно иметь в виду, однако, что в реальном эксперименте все четыре значения углов, рассчитанные с помощью формул (4)–(6), могут оказаться разными из-за шумов в измерительных приборах и в самом лазере. Поэтому с практической точки зрения корректное значение $\theta_i$ соответствует поиску минимального элемента последовательности { $\theta_1 - \varphi_1$, $\theta_1 - \varphi_2$, $\theta_2 - \varphi_1$, $\theta_2 - \varphi_2$ }. Так, если минимальное значение принимает элемент $\theta_2 - \varphi_1$, то разность фаз между импульсами следует положить равной $\theta_2$.

Значения разности фаз между соответствующими парами импульсов, определенные описанным только что способом, указаны рядом с соответствующими значениями нормированных сигналов на Рис. 4(г) слева. Поскольку описанный способ позволяет Алисе различать состояния с разностями фаз $\pi/2$ и $3\pi/2$, то она может использовать два неортогональных «фазовых» базиса: $X$-базис, использующий разности фаз 0 и $\pi$ для состояний $X_1$ и $X_0$, и $Y$-базис, использующий разность фаз $\pi/2$ и $3\pi/2$ для состояний $Y_1$ и $Y_0$. Однако при таком выборе углов большинство состояний Алисе придется отбросить, поэтому целесообразно использовать в качестве сигнального созвездия не точки на комплексной плоскости, а дуги, как это показано на Рис. 4(г) справа. Здесь для определенности мы выбрали центральные углы всех дуг равными $40°$. На Рис. 4(г) слева показано, что при таком выборе диапазона углов трем из восьми фреймов можно приписать определенные состояния: $Y_0$ для фреймов ##2,6 и $X_0$ для фрейма #1.

## XI. ОБУЖДЕНИЕ

Стоит отметить, что временная задержка между парами импульсов на Рис. 4(б), вообще говоря, не является обязательной. Более того, в отсутствии такой задержки автоматически увеличится частота приготовления состояний. Интерференционная картина, однако, в этом случае изменится и будет представлять собой последовательность импульсов со случайными амплитудами, которую следует разбить на фреймы по два временных слота (ранний и поздний) и работать только с четными или нечетными (в зависимости от начала отсчета) импульсами. С экспериментальной точки зрения, такая ситуация гораздо менее удобна, чем изображенная на рисунке Рис. 4(в), где центральные временные слоты всегда разделены импульсами с фиксированной амплитудой. Действительно, последние удобно использовать для нормировки сигнала, которая необходима для определения разности фаз по формулам (4)–(6). Кроме того, в отсутствии временной задержки между соседними состояниями разности фаз между импульсами оказываются связанными, так как в интерференции участвуют импульсы из соседних состояний. Строго говоря, в этом случае нельзя говорить о рандомизации фазы, поскольку Ева, вероятно, может получить информацию о фазах соседних состояний. Хотя подробно нами этот вопрос и не исследовался, но на первый взгляд кажется, что в этом случае Ева может попытаться провести атаку, аналогичную последовательной атаке (sequential attack [7]) на протокол КРК с дифференциальным фазовым кодированием [8]. По этим причинам использование временной задержки между соседними парами импульсов нам представляется целесообразным.

Другая особенность предлагаемой нами схемы заключается в использовании оптически связанных лазеров для формирования «пустых» временных слотов. Вообще говоря, импульсную последовательность на рисунке Рис. 4(б) можно создать с помощью одного лазера, используя для накачки соответствующий электрический сигнал. Однако, в этом случае форма соседних импульсов может существенно отличаться, особенно если ток смещения существенно ниже порога [6]. Это может отрицательно сказаться на видности интерференции, что приведет к росту уровня ошибок при распределении ключей, а также не позволит Алисе с высокой точностью определять разность фаз между импульсами.

Различие импульсов по форме в нерегулярной импульсной последовательности связанно с конечным временем жизни носителей, а точнее с тем, что эволюция числа носителей между импульсами, принадлежащими одной паре, будет отличаться от эволюции $N$ между импульсами, принадлежащими разным парам. Для уменьшения этого эффекта ток смещения обычно устанавливают над порогом. Такое решение, однако, не годится для рассматриваемого метода пассивного приготовления состояний, поскольку оно нарушает рандомизацию фаз импульсов [9]. Предлагаемый нами метод оптической инжекции с последующей оптической фильтрацией, как демонстрируют наши симуляции, не приводит к искажению формы импульсов, поскольку лазеры генерируют регулярные импульсные последовательности.

Отметим также возможность использования другого способа управления лазерами на Рис. 1, который может дополнительно повысить видность интерференции и уменьшить влияние джиттера. Мастер может создавать импульсную последовательность, аналогичную показанной на Рис. 4(б), т.е. генерировать пары импульсов с временной задержкой $2\Delta T$ между парами и накрывать, таким образом, 2/3 импульсов ведомого лазера. Центральная длина волны пропускания оптического фильтра в этом случае должна совпадать с длиной волны мастера, что позволит отфильтровать каждый третий импульс в сигнале ведомого лазера. В итоге мы вновь получим последовательность, показанную на Рис. 4(б), однако теперь оптические импульсы будут иметь меньший чирп и джиттер, что положительно скажется на их интерференции. Разумеется, как было сказано выше, импульсы мастера при такой накачке могут отличаться по форме, что вновь может отрицательно сказаться на интерференции импульсов ведомого лазера. Это особенно справедливо в случае, когда такой метод применяется на высокой частоте следования импульсов (более 1 ГГц). На невысоких частотах, однако, импульсы мастера можно сделать достаточно длинными, а импульсы ведомого лазера, напротив, короткими, что позволит нивелировать влияние формы импульсов мастера на форму импульсов ведомого лазера. Таким образом, этот альтернативный способ приготовления состояний может быть полезен при невысоких частотах следования импульсов.

## XII. ВЫВОДЫ

Предлагаемый нами метод пассивного приготовления квантовых состояний имеет высокий потенциал в контексте создания устройств для городских КРК сетей, хотя и имеет сегодня ограниченную практическую ценность ввиду необходимости использования интегрально-оптического фазового детектора. Стоимость последнего на сегодняшний день может превышать стоимость стандартных модуляторов на основе ниобата лития, что нивелирует преимущества данного подхода перед традиционными схемами КРК с активным приготовлением состояний по чисто экономическим причинам. Однако при дальнейшем развитии и удешевлении интегрально-оптических схем ценность нашего метода существенно возрастет. Кроме того, наш метод может применяться не только для фазовой, но и для амплитудно-фазовой модуляции, что существенно расширяет возможности приготовления квантовых состояний.